\documentclass[preprint,12pt,authoryear]{elsarticle}

\usepackage{amssymb}
\usepackage{amsmath}
\usepackage{hyperref}
\usepackage{url}
\usepackage{booktabs}   
\usepackage{tabularx}
\usepackage{amssymb}
\usepackage{graphicx}
\usepackage{subfig}
\usepackage{multirow}
\usepackage{siunitx} 
\usepackage{array}
\usepackage{enumitem}
\journal{Nuclear Physics B}

\begin{document}

\begin{frontmatter}

%% Title, authors and addresses

%% use the tnoteref command within \title for footnotes;
%% use the tnotetext command for theassociated footnote;
%% use the fnref command within \author or \affiliation for footnotes;
%% use the fntext command for theassociated footnote;
%% use the corref command within \author for corresponding author footnotes;
%% use the cortext command for theassociated footnote;
%% use the ead command for the email address,
%% and the form \ead[url] for the home page:
%% \title{Title\tnoteref{label1}}
%% \tnotetext[label1]{}
%% \author{Name\corref{cor1}\fnref{label2}}
%% \ead{email address}
%% \ead[url]{home page}
%% \fntext[label2]{}
%% \cortext[cor1]{}
%% \affiliation{organization={},
%%            addressline={}, 
%%            city={},
%%            postcode={}, 
%%            state={},
%%            country={}}
%% \fntext[label3]{}

\title{LLM-Driven Data Generation and a Novel Soft Metric for Evaluating Text-to-SQL in Aviation MRO} %% Article title

%% use optional labels to link authors explicitly to addresses:
%% \author[label1,label2]{}
%% \affiliation[label1]{organization={},
%%             addressline={},
%%             city={},
%%             postcode={},
%%             state={},
%%             country={}}
%%
%% \affiliation[label2]{organization={},
%%             addressline={},
%%             city={},
%%             postcode={},
%%             state={},
%%             country={}}

\author[inst1,inst2]{Patrick Sutanto}
\author[inst1,inst2]{Jonathan Kenrick}
\author[inst2]{Max Lorenz}
\author[inst1]{Joan Santoso}

% \affiliation[inst1]{organization={Soji.ai}}
% \affiliation[inst2]{organization={ Sains dan Teknologi Terpadu Surabaya (ISTTS)}}

% Define the affiliations using the labels
\address[inst1]{Institut Sains dan Teknologi Terpadu Surabaya (ISTTS)}
\address[inst2]{Soji.ai}

%% Abstract
\begin{abstract}
%% Text of abstract
The application of Large Language Models (LLMs) to text-to-SQL tasks promises to democratize data access, particularly in critical industries like aviation Maintenance, Repair, and Operation (MRO). However, progress is hindered by two key challenges: the rigidity of conventional evaluation metrics such as execution accuracy, which offer coarse, binary feedback, and the scarcity of domain-specific evaluation datasets. This paper addresses these gaps. To enable more nuanced assessment, we introduce a novel F1-score-based 'soft' metric that quantifies the informational overlap between generated and ground-truth SQL results. To address data scarcity, we propose an LLM-driven pipeline that synthesizes realistic question-SQL pairs from database schemas. We demonstrate our contributions through an empirical evaluation on an authentic MRO database. Our experiments show that the proposed soft metric provides more insightful performance analysis than strict accuracy, and our data generation technique is effective in creating a domain-specific benchmark. Together, these contributions offer a robust framework for evaluating and advancing text-to-SQL systems in specialized environments.

\end{abstract}

% %Graphical abstract
% \begin{graphicalabstract}
% %\includegraphics{grabs}
% \end{graphicalabstract}

% %Research highlights
% \begin{highlights}
% \item Introduces a novel F1-score-based 'soft' metric for text-to-SQL, focusing on practical utility over strict accuracy. 
% \item Proposes an LLM-driven methodology to generate text-to-SQL evaluation datasets, overcoming data scarcity. 
% \item Addresses critical evaluation challenges for text-to-SQL in specialized domains of aviation MRO.
% \end{highlights}

%% Keywords
\begin{keyword}
%% keywords here, in the form: keyword \sep keyword

%% PACS codes here, in the form: \PACS code \sep code

%% MSC codes here, in the form: \MSC code \sep code
%% or \MSC[2008] code \sep code (2000 is the default)

Large Language Models \sep Text-to-SQL \sep Evaluation Metrics \sep F1 Score \sep Aviation MRO \sep Data Generation.

\end{keyword}

\end{frontmatter}

\section{Background}
\label{background}

Recent breakthroughs have seen Large Language Models (LLMs) increasingly applied in diverse real-world scenarios \citet{achiam2023gpt, brown2020language, ouyang2022training, gemmateam2024gemmaopenmodelsbased, zhu2023minigpt}. However, LLMs are prone to issues like hallucination \citet{huang2025trusttrustenhancinglarge, mundler2023self, sun2024redeep}. Consequently, effectively evaluating their performance remains a significant challenge and an active research area \citet{panickssery2024llm, chang2024agentboard}, which can hinder their reliable application in critical domains.

A key application of LLMs is enabling natural language interaction with databases through text-to-SQL generation \citet{li2023llm}. Robust performance in text-to-SQL is vital to mitigate errors like hallucinations and ensure reliable outcomes in crucial real-world tasks. Although existing approaches have explored evaluation metrics for this task \citet{li2023llm, yu2018spider, qin2022survey}, we argue that current metrics are often overly strict. This rigidity makes it challenging to accurately assess performance, particularly with the limited evaluation datasets common in many practical settings. These challenges are particularly acute in high-stakes industries such as aviation.

The aviation industry, a vital driver of economic growth and global connectivity \citet{SUN2024100013, zhou2024competition}, prioritizes safety above all else \citet{feng2005analyzing}. Within this sector, effective Maintenance, Repair, and Operation (MRO) are paramount for ensuring in-flight safety and operational integrity \citet{van2013aircraft}. Enhancing the efficiency of MRO facilities can thus have a substantial positive impact \citet{vieira2016maintenance}. It is in this critical MRO context that text-to-SQL emerges as a promising application. Integrating text-to-SQL into LLM-powered chatbots would enable users to query databases using natural language, potentially yielding significant efficiency gains in MRO processes \citet{avers2012technical}.

However, despite its potential, deploying text-to-SQL effectively in the MRO domain faces significant hurdles. While numerous studies have explored LLMs for text-to-SQL generation, the MRO domain itself has received limited specific attention. This is compounded by the difficulty in acquiring suitable, domain-specific datasets. Furthermore, existing evaluation methods may not adequately address the unique operational requirements and safety considerations inherent in MRO. Collectively, these obstacles hinder efforts to ensure the reliability of LLM-generated SQL and to advance model capabilities within MRO. This is particularly problematic as most techniques for enhancing LLM performance depend heavily on high-quality data and dependable evaluation feedback \citet{pourreza2024chase, ramnath2025systematic}.

This paper addresses these critical gaps in MRO text-to-SQL evaluation. To overcome data scarcity, we introduce an LLM-driven data generation method that first synthesizes SQL queries from table schemas and subsequently derives corresponding natural language questions. Furthermore, recognizing that standard metrics like execution accuracy often inadequately capture the practical outcomes when LLMs process query outputs, we propose a novel F1-score-based 'soft' metric. This new metric is designed to assess the utility of the retrieved information more holistically. The combined value of our data generation technique and this novel metric is demonstrated through an empirical evaluation of diverse LLMs on an authentic MRO database.

\section{Method details}
\label{method_details}
Our methodology is structured in four parts. First, we present the metrics for evaluating the performance of text-to-SQL model. Second, we describe the dataset creation process. Third, we detail the text-to-SQL system targeted for evaluation. Fourth, we outline the metrics used to analyze the characteristics of the generated dataset.

\subsection{Model Evaluations Metrics}
\label{subsec:model_eval_metrics}

To evaluate text-to-SQL model performance, we employ three distinct metrics: (1) Execution Accuracy, which determines if the model-generated SQL's execution results perfectly match those of the ground-truth SQL; (2) Execution Error Count, quantifying instances where model-generated SQL fails to execute; and (3) our novel Soft Execution F1 score, designed to measure the informational overlap between the predicted and ground-truth SQL execution results.

\subsubsection{Execution Accuracy}

Execution Accuracy is determined through a multi-step comparison of the execution results from the predicted SQL and the ground-truth SQL.

First, we check if the dimensions (number of rows and columns) of the two result sets are identical. Any discrepancy at this stage immediately marks the prediction as incorrect.
If the dimensions match, we then attempt to establish a one-to-one correspondence between columns in the predicted result set and columns in the ground-truth result set. A predicted column is considered matched to a ground-truth column if all their respective values are identical and appear in the same order row-wise.

The prediction is deemed correct if and only if every column in the ground-truth result set finds such a unique, content-identical match in the predicted result set, and all columns in the predicted set are similarly matched. Otherwise, the prediction is marked incorrect.

\subsubsection{Soft Execution F1}
We observe that traditional execution accuracy often yields overly strict evaluations. For instance, it would typically penalize a predicted query like \verb|SELECT * FROM table LIMIT 4| as entirely incorrect if the ground truth were \verb|SELECT * FROM table LIMIT 5|, despite the substantial overlap in retrieved information. However, LLMs tasked with generating natural language answers from SQL query results can often tolerate such minor discrepancies, still producing useful output even if the underlying SQL is not an exact match to a predefined ground truth. As this scoring function is defined for a data point, we averaged the results over all data point in the dataset.

We assume that an execution result is a matrix, where the columns represent each attribute and the rows represent each value. Our proposed metrics compare ground truth SQL results $G \in R^{M \times N}$ to the predicted SQL execution results $P \in R^{S \times T}$. We first compute the pairwise score between each column of $G$ and $P$. For each column of the ground truth column, denoted $G_{:,n} \in R^M$, we compute the precision, recall, and F1 score when comparing it with each of the predicted columns denoted $P_{:,t} \in R^S$. The computation can be done as shown in the formula below.

$$
Precision_{n,t} =  \frac{1}{S}\sum_{s=1}^S I[P_{s,t} \in G_{:,n}]
$$

$$
Recall_{n,t} =  \frac{1}{M}\sum_{m=1}^M I[G_{m,n} \in P_{:,t}]]
$$

$$
F1_{n,t} =  \frac{2 \times Precision_{n,t} \times Recall_{n,t}}{Precision_{n,t} + Recall_{n,t}}
$$

Where $I[P_{s,t} \in G_{:,n}]$ denote indicator function that take value one when the prediction value exist in the n'th column of the ground truth, and 0 otherwise, whereas $I[G_{m,n} \in P_{:,t}]]$ is defined similarly. After computing each of the columns' pairwise scores, we take the maximum value as the score for that column. We then compute precision, recall, and F1-score for the column level, and use the F1-score as the final score. This process results in a score between 0-1. 
$$
Precision =  \frac{1}{T}\sum_{t=1}^T max(F1_{:,t})
$$

$$
Recall =  \frac{1}{N}\sum_{n=1}^N max(F1_{n,:})
$$

$$
F1 =  \frac{2 \times Precision \times Recall}{Precision + Recall}
$$

The core intuition for our proposed metric is to quantify the similarity between the information retrieved from executing the ground truth SQL and that from the predicted SQL. A key requirement for such a metric is the ability to effectively handle instances where the query results differ in their row or column counts. To address this, we employ an F1-inspired approach for comparing the content of individual columns and the overall structure of the result sets. This allows the metric to penalize predicted results that are either too sparse (i.e., missing relevant information) or too verbose (i.e., including excessive, irrelevant data) when compared to the ground truth. The rationale is that both missing and superfluous data can act as noise, potentially degrading the quality of answers generated by an LLM that subsequently processes these SQL outputs.

\subsubsection{Execution Error Count}

A common issue with Large Language Model (LLM)-generated SQL is that not all outputs are syntactically correct or semantically valid for execution against the target database. The Execution Error Count metric quantifies the number of LLM-generated SQL queries that fail to execute successfully. This metric serves as a indicator of the LLM's ability to produce usable SQL. A high error count suggests that the LLM struggles with basic SQL syntax, adherence to the database schema, or other constraints, rendering its outputs frequently unusable without manual intervention. Conversely, a low error count indicates greater reliability in producing queries that can at least be processed by the database system, a crucial first step before evaluating the correctness of the retrieved results.

\subsection{Dataset Generation Process}
\label{subsec:dataset_generation}

The database we use contains information about tickets and aircraft layovers. Our objective is to create an evaluation dataset for text-to-SQL systems designed to query this database. To achieve this, we leverage the commercial Large Language Model (LLM) gemini-2.0-flash \citet{team2023gemini}. This model was chosen for its balance of performance and cost, and its ability to process very long contexts, a crucial capability for our use case. The data generation follows a two-stage process: the first stage involves using the LLM to generate SQL queries and their corresponding natural language questions based on the database schema and sample data. The second stage comprises rigorous filtering of these generated pairs to ensure high data quality for the evaluation dataset.

\subsubsection{Question Generation}
The prompt used for generating question-SQL pairs is detailed in Appendix \ref{appendix:dataset_generation_prompt}. The database utilized contains only two tables: one for tickets and one for aircraft layovers. For each table, we crafted a textual description detailing its column names and their respective meanings. This schema information, along with a few randomly sampled rows from each table (to enhance output diversity), was incorporated into the main generation prompt. Our approach reverses the conventional process of generating an SQL query from a given question. Specifically, we instructed the LLM to first generate an SQL query and then formulate a corresponding natural language question based on that query. This SQL-first methodology aims to ensure that generated questions are directly answerable by their paired SQL queries, thereby enhancing their relevance and utility for evaluation.

To ensure the generated data was structured and easily parsable, we included specific instructions in the prompt for the LLM. The model was instructed to provide its output in JSON format. This JSON output was structured as a list of dictionaries, each encapsulating a generated SQL query and its corresponding natural language question, reinforcing the desired SQL-first output sequence. Requesting multiple question-SQL pairs in a single API call also offered computational and cost efficiencies compared to generating them individually. To further promote output diversity, a high temperature setting of 1.5 was employed during generation.

To enhance both the quality and diversity of the generated outputs, we instructed the model to first articulate its reasoning process as a string before generating the final SQL-question pair. Incorporating such a reasoning step, akin to chain-of-thought prompting, is known to improve LLM performance across various tasks \citet{kojima2022large}. The prompt also explicitly encouraged creativity from the LLM to further diversify both the generated reasoning and the resulting SQL-question pairs. To better mimic real-world scenarios where users might pose simple-sounding questions that necessitate complex underlying queries, we also instructed the LLM to occasionally generate intricate SQL queries paired with concise natural language questions. This instruction was intended to ensure a wider range of query complexities in the dataset and create more realistic, challenging evaluation scenarios for text-to-SQL models.

\subsubsection{Answer Generation and Filtering Process}
To enhance the quality of the evaluation dataset, we implemented several filtering steps. Initially, each generated SQL query was validated for executability against the database; non-executable queries were discarded. Subsequently, queries that executed successfully but returned an excessive number of results (e.g., exceeding a predefined threshold) were also filtered out. This process ensured that the final dataset comprised only valid SQL queries yielding manageable result sets suitable for downstream answer generation.

Since a primary goal of text-to-SQL systems is to ultimately provide a natural language answer to a user's question, we proceeded to generate such answers for the filtered SQL-question pairs. The LLM prompt used to generate a human-readable natural language answer from the executed SQL results is detailed in Appendix \ref{appendix:answer_generation_prompt}. This prompt instructed the LLM, similar to the initial data generation, to articulate its reasoning before formulating the final answer. Crucially, the LLM was also tasked with assessing the sufficiency of the SQL query's results for answering the question; pairs deemed insufficient by the LLM were filtered out. This served as a further quality control measure, ensuring that each question in the dataset was clearly answerable by its corresponding ground-truth SQL execution results. Combined, these generation and multi-stage filtering processes yield a high-quality dataset of question-SQL-answer triplets, suitable for robust text-to-SQL evaluation and potentially for future model development.

\subsection{Text-to-SQL System}
\label{subsec:text_to_sql}
For our text-to-SQL system evaluations, we utilized several publicly available Large Language Models (LLMs). The input prompts provided to these LLMs included comprehensive database schema information (table names, column names, data types, and descriptions), details of table relationships, and task-specific instructions for generating SQL queries. Consistent with our data generation phase, the LLMs were instructed to output their reasoning process first, followed by the SQL query, all encapsulated within a JSON object. The models evaluated include specific versions from four open-source model families: DeepSeek-V3, Mistral-Small-24B-Instruct-2501, Phi-4, and Llama-3.3-70B-Instruct. For each model, we evaluated performance under two distinct prompting strategies: (1) a zero-shot setting, where prompts contained only task instructions and schema information; and (2) a few-shot setting, where prompts were augmented with four illustrative examples. These examples showcased the desired input-to-output transformation, complete with reasoning steps and the SQL query in the specified JSON format.

\subsection{Dataset Evaluations Metrics}
\label{subsec:dataset_eval_metrics}
Analysis of the dataset generated by the Gemini model involves several approaches. We first assess the difficulty of the generated questions by measuring the complexity of their corresponding SQL queries. Subsequently, we evaluate dataset diversity and the extent of duplication using exact match counts and maximum semantic similarity metrics.

\subsubsection{SQL Difficulty Measurement}
\label{subsubsec:sql_difficulty_measure}

Drawing inspiration from methodologies like those used for the Spider dataset \citet{yu2018spider}, we estimate SQL query difficulty. Query difficulty is classified into four categories: easy, medium, hard, and extra. This estimation involves assessing three types of SQL components, where the presence of specific keywords within a query serves as an indicator variable:
\begin{itemize}
  \item Component 1 Keywords: WHERE, GROUP BY, ORDER BY, LIMIT, JOIN, OR, LIKE.
  \item Component 2 Keywords: EXCEPT, UNION, INTERSECT.
  \item Other Component Keywords: Multiple instances of SELECT, WHERE, or GROUP BY.
\end{itemize}
We sum the indicator variables for each of these component types, and then determine the overall query hardness based on the conditions outlined in Table \ref{tab:hardness_conditions}.

\begin{table}
\centering
\begin{tabular}{ l|c|c|c|l } 
\hline
Difficulty & comp. 1 & comp. 2  & Other & Notes \\
\hline
Easy &  $\leq 1$ & $=0$  & $=0$  & All conditions must be met \\ 
\hline
\multirow{2}{4em}{Medium} & $\leq 1$ & $\leq 2$ & $=0$ & \multirow{2}{12em}{Satisfies either of this rules} \\ 
& $\leq 2$ & $\leq 1$ & $= 0$ & \\ 
\hline
\multirow{3}{4em}{Hard} & $\leq 2$ & $> 2$ & $=0$ & \multirow{3}{12em}{Satisfies either of this rules} \\ 
& $ > 2 $ AND $ \leq 3 $ & $\leq 2$ & $= 0$ & \\ 
& $\leq 1$ & $=0$ &   $\leq 1$ & \\ 
\hline
Extra &  Other & Other  & Other  & Other conditions \\

\hline
\end{tabular}
\caption{SQL Query Difficulty Classification Rules}
\label{tab:hardness_conditions}
\end{table}

\subsubsection{Exact Match Percentage}
Our dataset generation process yields triplets consisting of a natural language question, its corresponding SQL query, and an LLM-generated natural language answer. To assess internal duplication within this dataset, we measure the frequency of exact matches. This analysis is performed independently for the sets of questions, SQL queries, and natural language answers. An exact match is defined as two strings being identical after both are converted to lowercase (case normalization).

\subsubsection{Average Maximum Semantic Similarity}
To further assess dataset diversity beyond exact matches, we analyze semantic similarity. The generated questions, SQL queries, and natural language answers are independently embedded into a dense vector space using a pre-trained model from the Sentence-Transformers library \citet{reimers-2019-sentence-bert} (specifically, all-MiniLM-L6-v2). The semantic similarity between any two resulting vector embeddings, x and y, is then computed using cosine similarity

$$sim(x,y) = \frac{x \cdot y}{||x||^2_2 \cdot ||y||^2_2}$$

For each item (i.e., each embedded question, SQL query, or answer), we compute its cosine similarity with all other items of the same type within the dataset. We then identify its maximum similarity score to any other item of that type (excluding self-similarity). The 'Average Maximum Semantic Similarity' is calculated by averaging these maximum scores, performed separately for questions, SQL queries, and answers. This metric serves as an indicator of the overall semantic coherence or internal similarity within each category of generated data. We anticipate this metric will yield relatively high values given the specific MRO domain, where generated items inherently share considerable thematic content.

\section{Results and Analysis}
This section presents our findings, divided into two main parts: Dataset Analysis and Model Evaluation Analysis.

\subsection{Dataset Analysis}
In this subsection, we analyze the characteristics of the generated evaluation dataset. Our analysis begins with an overview of key dataset statistics. We then investigate the extent of data duplication (both exact matches and semantic similarity) and conclude by examining the difficulty distribution of the SQL queries within the dataset.

\subsubsection{Dataset Statistics}
To facilitate text-to-SQL evaluation, we generated a dataset comprising 1079 question-SQL pairs. The statistical properties of this dataset are now examined, beginning with the usage patterns of different SQL syntax elements. Specifically, we analyzed the prevalence of constructs like GROUP BY, HAVING, and JOIN. The findings from this syntax analysis are summarized in Table \ref{tab:dataset_statistics}.

\begin{table}
\centering
\begin{tabular}{ l|c|c } 
\hline
Type &  Count & Percentage \\ 
\hline
GROUP BY &  775 & 71.83 \\ 
ORDER BY &  519 & 48.1 \\ 
JOIN & 222 & 20.57 \\
Sub-Query &  159  & 14.74 \\ 
HAVING &  136 & 12.60 \\
DISTINCT &  111 & 10.29 \\ 
CASE &  100 & 9.27 \\ 
WITH &  22 & 2.04 \\ 

\hline
\end{tabular}
\caption{Dataset Statistics}
\label{tab:dataset_statistics}
\end{table}

The SQL queries generated by Gemini exhibit considerable complexity. For instance, over 70\% of these queries include GROUP BY clauses. Furthermore, subqueries are present in more than 10\% of the instances, and CASE statements were utilized in over 9\% of the generated queries. The prevalence of these advanced SQL features demonstrates Gemini's capability to construct intricate and challenging SQL queries, contributing to a robust dataset for evaluation.

\subsubsection{SQL Difficulty}
We further analyzed the difficulty distribution of the SQL queries generated by the Gemini model, based on the classification scheme detailed in Section \ref{subsubsec:sql_difficulty_measure}. Table \ref{tab:dataset_diff_stats} presents the count and percentage of queries falling into each difficulty category of easy, medium, hard, and extra hard.

\begin{table}
\centering
\begin{tabular}{ l|c|c } 
\hline
Type &  Count & Percentage \\ 
\hline
Easy &  301 & 27.9 \\ 
Medium &  301 & 27.9 \\ 
Hard & 244 & 22.6 \\
Extra &  233  & 21.6 \\ 
\hline
\end{tabular}
\caption{Dataset Difficulty Distributions}
\label{tab:dataset_diff_stats}
\end{table}

The results in Table \ref{tab:dataset_diff_stats} indicate that the queries generated by Gemini are distributed relatively evenly across the defined difficulty levels, approaching a uniform distribution. This balance suggests Gemini's capability to produce a dataset with a good spectrum of query complexities, making it well-suited for evaluating text-to-SQL model performance across varying levels of challenge.

\subsubsection{Similarity Between Data}
\label{susubsec:data_similarity}

To assess the internal diversity of our generated dataset, we analyzed the uniqueness of its primary components: the natural language questions, the SQL queries, and the corresponding textual answers. Specifically, we identified the extent of exact matches for each of these components across the entire dataset. The counts and percentages of these exact matches are detailed in Table \ref{tab:data_similarity}.

\begin{table}
\centering
\begin{tabular}{ l|c|c|c } 
\hline
Type &  Match Count & Match \% & Maximum Similarity \\ 
\hline
Question &  43 & 3.99 & 0.84 $\pm$ 0.12  \\ 
SQL Query &  519 & 48.1 & 0.93 $\pm$ 0.06\\ 
Answer & 222 & 20.57 & 0.85 $\pm$ 0.11 \\

\hline
\end{tabular}
\caption{Dataset Statistics}
\label{tab:data_similarity}
\end{table}

The analysis revealed a notably high percentage of duplicate SQL queries within the dataset. However, given that the corresponding natural language questions and generated answers exhibited significantly less duplication, we opted not to filter out these repeated SQL queries. We posit that retaining these instances is valuable for assessing a model's robustness to diverse phrasings of questions that map to the same underlying SQL logic.

For semantic similarity assessment, we employed the all-MiniLM-L6-v2 model from the sentence-transformers library to generate embeddings for the questions, SQL queries, and answers. From these embeddings, we calculated cosine similarity scores. The average maximum semantic similarity scores were relatively high across all components. This outcome is largely expected, given the dataset's specific focus on the aviation MRO domain, where a degree of topical overlap is inherent. Nevertheless, despite these high average scores reflecting domain specificity, the observed variations suggest that the dataset possesses sufficient diversity for our initial evaluation purposes.

\subsection{Model Evaluation Analysis}

Our model evaluation analysis assesses the performance of various LLM backbones under both few-shot and zero-shot prompting strategies. We then examine their performance across the different SQL difficulty categories within our dataset.

Given the presence of duplicate SQL queries in our evaluation dataset, as noted in Section \ref{susubsec:data_similarity}, we employ a specific aggregation method for reporting performance metrics. To ensure a more representative evaluation, particularly when considering varied question phrasings for the same underlying SQL logic, we first average the scores for all instances sharing an identical ground truth SQL query. The final reported metrics are then averaged across these unique SQL query groups. This two-step averaging approach is based on the premise that identical SQL queries often correspond to questions with similar intent but different phrasings. By first averaging scores for each unique SQL query, we aim to provide a more robust estimation of a model's performance on distinct SQL tasks, thereby better reflecting its capability across diverse questions.

\subsubsection{Baseline Evaluation}
For our comparative analysis, we evaluated the text-to-SQL performance of four Large Language Models (LLMs): DeepSeek-V3, Llama-3-70B-Instruct, microsoft/phi-4, and Mistral-Small-24B-Instruct. Each of these models was assessed under two distinct prompting conditions: zero-shot and few-shot. The evaluation relied on three primary metrics: our novel Soft Execution F1 score, the conventional Execution Accuracy, and the Execution Error Count.

\begin{table}
\centering
  \begin{tabular}{lSSSSSS}
    \toprule
    \multirow{2}{*}{Model} &
      \multicolumn{2}{c}{Soft Execution F1} &
      \multicolumn{2}{c}{Execution Accuracy} &
      \multicolumn{2}{c}{Error Count} \\
      \cmidrule(lr){2-3} \cmidrule(lr){4-5} \cmidrule(lr){6-7}
      & {0-shot} & {few-shot} & {0-shot} & {few-shot} & {0-shot} & {few-shot} \\
      \midrule
    DeepSeek & 81.5 & 82.1 & 24.1 & 23.7 & 61.5 & 58.9 \\
    Phi-4 & 78.9 & 78.6 & 21.9 & 19.4 & 72.6 & 73.6 \\
    Llama & 78.5 & 77.9 & 20.7 & 20.8 & 66.0 & 69.0 \\
    Mistral & 73.9 & 68.8 & 20.1 & 18.0 & 81.7 & 108.3 \\
    \bottomrule
  \end{tabular}
  \caption{Model Performance with 0-shot and few-shot settings}
  \label{table:model_performance}
\end{table}

A primary observation from our evaluations is that the DeepSeek-V3 model significantly outperformed the other models and consistently exhibited the best results across all metrics. Furthermore, we noted that in several instances, employing few-shot prompting paradoxically led to a degradation in performance compared to zero-shot prompting. Several factors might explain this counterintuitive outcome. One hypothesis is that the provided few-shot examples, while intended to guide, may inadvertently introduce noise or overly constrain the model's reasoning, leading it to mimic the examples rather than generalizing effectively. Conversely, zero-shot prompting might allow certain models, particularly those less adept at in-context learning from limited examples, to leverage their broader pre-trained knowledge more effectively. Lastly, a consistent trend indicated that models achieving higher overall accuracy (on both Execution Accuracy and Soft Execution F1 metrics) generally also had lower Execution Error Counts, suggesting a correlation between the ability to generate semantically correct SQL and syntactically valid SQL.

An interesting finding further illustrates why strict Execution Accuracy can sometimes be misleading when evaluating practical LLM performance. For instance, our results indicate that Zero-Shot Mistral-Small achieved a higher Execution Accuracy than Few-Shot Phi-4. However, under the same comparison, Phi-4's Soft Execution F1 score was significantly higher than that of Mistral-Small. Upon closer inspection, we found numerous instances where Phi-4's generated SQL, while not perfectly matching the ground truth, retrieved highly relevant information but was nonetheless marked as entirely incorrect by the strict Execution Accuracy metric. For example, in response to a question asking for the MRO shop with the most tickets, Phi-4's query correctly identified the shop but omitted the associated ticket count, whereas the ground truth SQL included both. In this scenario, our Soft Execution F1 score appropriately assigned partial credit (approximately 0.666), reflecting the partial correctness, while Execution Accuracy yielded a score of 0. Compounding this, the Execution Error Count for Mistral-Small was also considerably higher than for Phi-4 in these settings. This case underscores the need for more fine-grained metrics like Soft Execution F1, which better reflect the practical utility of an LLM's output when the primary goal is to answer user questions effectively.

\subsubsection{Evaluation Based on Category}

To analyze how performance varies with SQL query difficulty, we calculated the average scores for our across all evaluated models and both their zero-shot and few-shot prompting configurations for each distinct difficulty category.This explicitly key evaluation metrics (Execution Accuracy, Soft Execution F1, and Execution Error Count) across all models and These aggregated performance figures are summarized in Table \ref{tab:difiiculty_score}.

\begin{table}
\centering
\begin{tabular}{ l|c|c|c } 
\hline
Type &  Soft Execution F1 & Execution Accuracy & Error Count  \\ 
\hline
Easy &  84.8 & 18.9 & 7.2 \\ 
Medium &  75.5 & 12.9 & 21.8 \\ 
Hard & 77.8 & 23.8 & 21.6 \\
Extra &  72.4  & 30.2 & 23.4 \\ 
\hline
\end{tabular}
\caption{Dataset Difficulty Distributions}
\label{tab:difiiculty_score}
\end{table}

Our evaluation across different difficulty levels revealed a somewhat counterintuitive finding: on average, models performed notably better on 'hard' questions compared to 'medium' ones, although overall performance remained lowest for the 'extra hard' category. Furthermore, when specifically considering the Execution Accuracy metric, performance on 'easy' questions was surprisingly lower than on both 'hard' and even 'extra hard' questions. We hypothesize that this anomaly arises because 'easy' questions often permit multiple valid SQL formulations to retrieve the same or equivalent information. Consequently, the strict nature of Execution Accuracy penalizes valid alternative solutions if they do not exactly match the single ground truth query, even if the retrieved data is functionally correct for the user's intent. This observation further suggests that relying solely on Execution Accuracy can lead to less intuitive performance assessments, particularly when contrasted with metrics like our Soft Execution F1, which are designed to be more tolerant of such functionally equivalent variations.

In addition to the aggregated overview, we provide a more granular analysis of each model's performance across the different SQL difficulty categories. For this presentation, the scores for each model within a specific difficulty level are an average of its performance under both zero-shot and few-shot prompting conditions. This averaging is based on the rationale that a model demonstrating strong, consistent performance across both settings can be considered more robust and versatile. These detailed per-model results are presented in Table \ref{tab:model_difiiculty_score} below.

\begin{table}
\centering
\begin{tabular}{ l|c|c|c|c|c } 
\hline
Model &  Easy & Medium & Hard & Extra  \\ 
\hline
DeepSeek &  87.8 & 79.8 & 82.0 & 77.9 \\ 
Phi-4 & 84.9 & 76.1 & 82.6 & 72.0 \\
Llama &  84.8 & 77.7 & 77.2 & 73.0 \\ 
Mistral &  81.6  & 68.3 & 69.4 & 66.7 \\ 
\hline
\end{tabular}
\caption{Dataset Difficulty Distributions}
\label{tab:model_difiiculty_score}
\end{table}

Consistent with earlier observations, the DeepSeek-V3 model generally outperformed the other models across most difficulty categories. More surprisingly, on this MRO-specific text-to-SQL task, Phi-4 often performed better than, or at least comparably to, Llama-3-70B-Instruct. We hypothesize this unexpected outcome may be attributed to Phi-4's potentially stronger grasp of the specialized aviation domain knowledge, coupled with its competent code generation capabilities which appeared competitive on this particular dataset. This hypothesis finds some support in Phi-4's technical report, which highlights its strong performance on complex reasoning and science-related benchmarks. Such capabilities might translate to a better understanding of domain-specific nuances in our MRO task, even when compared against larger models like Llama-3-70B-Instruct.

\bibliographystyle{elsarticle-harv}
\bibliography{references}

%% The Appendices part is started with the command \appendix;
%% appendix sections are then done as normal sections
\appendix
\section{Appendix}
\label{appendix}

\subsection{Prompts}
We include the prompt used for dataset generation in this sections. 
\subsubsection{Dataset Generation Prompt}
The prompt to generate the SQL and question is provided on Table \ref{table:json_few_shot_arc_easy}

\label{appendix:dataset_generation_prompt}

\begin{table*}
\small
\begin{tabular}{l}
\begin{tabularx}{\textwidth}{m{13.5cm}}
\toprule
\{AIRCRAFT\_LAYOVERS\_TABLE\_EXPLANATION\_PROMPT\} \\  
\{TICKET\_TABLE\_EXPLANATION\_PROMPT\} \\
\\
Example Layover Data \\ 
\{random aircraft layovers row examples\} \\ 
\\
Example Ticket data: \\ 
\{random ticket row examples\} \\ 
\\
Your task is to write a SQL query that able to answer a question. \\ 
First, you need to generate the sql query \\ 
Second, create question based on that sql query (what question is answered by the query) \\
\\
\# IMPORTANT NOTE \\
- Your answer must be in JSON format, a list of dictionaries containing \{\{"sql\_query": <str>, "question": <str>\}\} \\ 
- Ensure that you create reasoning on what kind of sql you need to create step-by-step Before answering in JSON!\\ 
- Create the reasoning first on top step-by-step, then generate all the JSON bellow it! \\
- The sql\_query string must be a valid SQL query that can be executed on the database! \\ 
- Be creative and Create diverse and complete reasoning, sql query and its question! \\ 
- Also try to create difficult SQL that have very short question (yo may also consider to use abbreviations only)! \\ 
- Include a very difficult SQL that you can think of! \\ 
- Reasoning should be as complete as possible!  \\ 

\bottomrule
\end{tabularx}
\end{tabular}
\vspace{2pt}
\caption{Dataset generation prompt.}
\label{table:json_few_shot_arc_easy}
\end{table*}

\subsubsection{Generate Answer Prompt}
\label{appendix:answer_generation_prompt}

The prompt to generate the answer and also for filtering is provided on Table \ref{table:answer_prompt}

\begin{table*}
\small
\begin{tabular}{l}
\begin{tabularx}{\textwidth}{m{13.5cm}}
\toprule

Answer the question based on retrieved context \\ 
\# Guidelines : \\ 
1. Think step-by-step before answering using JSON format. \\ 
2. Think whether the context is relevant or not, and how to answer the question using the retrieved context \\ 
3. Answer with JSON format of the following format: \\ 
\{ \\ 
    "reasoning": <str>, // Your thinking \\ 
    "context\_is\_sufficient" : <boolean>, // True if the context is sufficient to answer \\  question, False if not \\ 
    "answer": <str> // string in human-readable answer \\ 
\} \\ 
\\ 
\# SQL Query used for retrieval : <SQL\_QUERY> \\ 
\\ 
\# Retrieved Context \\ 
<CONTEXT> \\ 
\\ 
\# Question \\ 
<QUESTION>\\ 
\\ 
Answer in JSON format, and ensure that the "answer" part is human-readable: \\

\bottomrule

\end{tabularx}
\end{tabular}
\vspace{2pt}
\caption{Answer generation prompt.}
\label{table:answer_prompt}
\end{table*}

\subsection{Limitations}
Our proposed Soft Execution F1 metric, while offering flexibility, has certain limitations. Primarily, it is order-insensitive, meaning it evaluates column content based on the presence and frequency of values without regard to their row sequence. While often acceptable for set-based comparisons, this means it doesn't capture nuances where row order is critical. Another consideration involves its application to columns with a very small set of unique values, such as Boolean or low-cardinality categorical data. In many practical text-to-SQL applications, the focus is on retrieving the correct set of entities or values, where moderate insensitivity to order is often acceptable.

\subsection{Future Works}
Future research will pursue two primary directions: enhancing the quality of our LLM-generated evaluation dataset and further refining the proposed Soft Execution F1 metric.

Firstly, we aim to improve the fidelity and robustness of the evaluation where result order is semantically important. Further investigations could also explore mechanisms for differential weighting of columns based on their perceived importance or more nuanced dataset. While our current data generation approach provides a valuable starting point, we recognize that some generated ground truth instances (questions, SQL queries, or natural language answers) may occasionally be suboptimal. To address this, we plan to explore more sophisticated prompt engineering approaches to handling various data types within the comparison logic.

\end{document}